\documentclass[review,number,sort&compress,3p,times]{elsarticle}

\usepackage{hyperref}

\usepackage{lineno}
\usepackage{color}

\begin{document}

\begin{frontmatter}

\title{Study of Light Backgrounds from Relativistic Electrons in Air Light-Guides}


\author[mymainaddress,argonne]{S. Riordan\fnref{anlpres}}
\author[mymainaddress]{Y. X. Zhao\corref{mycorrespondingauthor}}
\cortext[mycorrespondingauthor]{Corresponding author}
\ead{yxzhao@jlab.org}

\author[KPH_Mainz]{S.~Baunack}
\author[KPH_Mainz]{D.~Becker}
\author[mymainaddress]{C.~Clarke}
\author[mymainaddress]{K.~Dehmelt}
\author[mymainaddress]{A.~Deshpande}
\author[manitoba]{M.~Gericke}
\author[KPH_Mainz]{B.~Gl\"aser}
\author[KPH_Mainz]{K.~Imai}
\author[mymainaddress]{T.~Kutz}
\author[KPH_Mainz,GSI_Darmstadt,HIM_Mainz]{F.~E.~Maas}
\author[idahostate]{D.~McNulty}
\author[manitoba]{J.~Pan}
\author[mymainaddress]{S.~Park}
\author[manitoba]{S.~Rahman}
\author[syracuse]{P.A.~Souder}
\author[manitoba]{P.~Wang}
\author[mymainaddress]{B.~Wellman}
\author[mymainaddress]{K.~S.~Kumar}

\address[mymainaddress]{Department of Physics and Astronomy, Stony Brook University, Stony Brook, NY 11794, USA}
\address[argonne]{Physics Division, Argonne National Laboratory, Argonne, IL 60439, USA}
\address[KPH_Mainz]{Institute for Nuclear Physics, Johannes Gutenberg-University Mainz, J. J. Becherweg 45, D-55099 Mainz, Germany}
\address[manitoba]{Department of Physics and Astronomy, University of Manitoba,
Winnipeg, MB R3T 2N2, Canada}
\address[GSI_Darmstadt]{GSI Helmholtzzentrum f\"ur Schwerionenforschung GmbH, Planckstra\ss{}e 1, D-64291 Darmstadt, Germany}
\address[HIM_Mainz]{Helmholtz-Institute Mainz, Johannes Gutenberg-University Mainz, D-55099 Mainz, Germany}
\address[idahostate]{Department of Physics, Idaho State University, Pocatello, ID 83209, USA}
\address[syracuse]{Department of Physics, Syracuse University, Syracuse, NY 13244, USA}

\fntext[anlpres]{Present address: Physics Division, Argonne National Laboratory, Argonne, IL 60439, USA}

\begin{abstract}
The MOLLER experiment proposed at the Thomas Jefferson National Accelerator Facility plans a precision low energy determination of the weak mixing angle via the measurement of the parity-violating asymmetry in the scattering of high energy longitudinally polarized electrons from electrons bound in a liquid hydrogen target (M{\o}ller scattering). A relative measure of the scattering rate is planned to be obtained by intercepting the M{\o}ller scattered electrons with a circular array of thin fused silica tiles attached to air light guides, which facilitate the transport of Cherenkov photons generated within the tiles to photomultiplier tubes (PMTs). The scattered flux will also pass through the light guides of downstream tiles, generating additional Cherenkov as well as scintillation light and is a potential background. In order to estimate the rate of these backgrounds, a gas-filled tube detector was designed and deployed in an electron beam at the MAMI facility at Johannes Gutenberg University, Mainz, Germany. Described in this paper is the design of a detector to measure separately the scintillation and Cherenkov responses of gas mixtures from relativistic electrons, the results of studies of several gas mixtures with comparisons to simulations, and conclusions about the implications for the design of the MOLLER detector apparatus. 
\end{abstract}


\end{frontmatter}


\section{Introduction}


The next generation of precision tests of the Standard Model at modern accelerator facilities requires the use of large area, open geometry detectors to obtain the required statistical precision, which in turn necessitates novel detector configurations and experimental techniques.  The MOLLER experiment~\cite{Benesch:2014bas,moller_homepage,moller_proposal} intends to study a parity-violating signal in the M{\o}ller scattering process to measure the weak mixing angle $\sin^2\theta_W$ at the Thomas Jefferson National Accelerator Facility. It will utilize several concentric rings of fused silica Cherenkov radiators as the primary detectors for the scattered electrons. Such an arrangement allows for in principle the total azimuthal coverage of the M{\o}ller scattering process, but due to the large acceptance one must carefully account for a number sources of background, including those from the primary flux itself. 

\begin{figure}
\begin{center}
\resizebox{0.95\textwidth}{!}{ \includegraphics{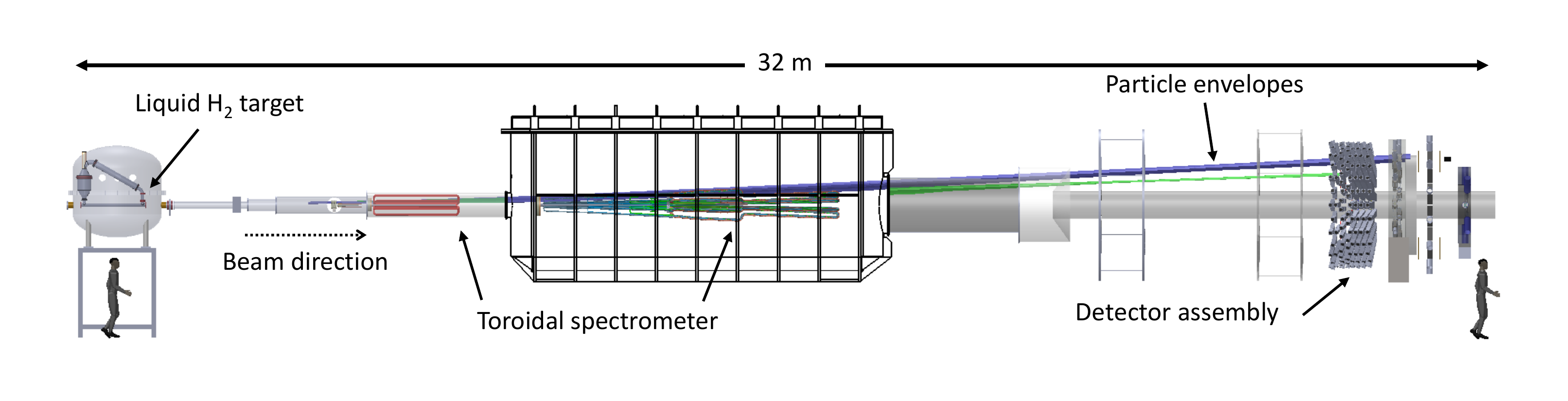} }
\caption{A schematic of the MOLLER experiment setup.}
\label{fig:MOLLER_overview}
\end{center}
\end{figure}

Shown in Figure \ref{fig:MOLLER_overview},
the overall experimental design calls for an 11~GeV polarized electron beam to impinge on a 1.5~m liquid hydrogen target with a spectrometer and collimator system designed to collect M{\o}ller scattered electrons in the full range of the azimuth, and a center-of-mass polar angle range of 60 to 120 degrees. This corresponds to laboratory scattering angles of 5~to 20~mrad.  A $1\pi$-collimation system with an odd number of apertures symmetric in center-of-mass polar angle is designed to detect one of the pair of indistinguishable electrons in the M{\o}ller scattering process, hence offering $2\pi$ azimuthal coverage as both electrons are in the same scattering plane.  Two toroidal magnetic elements will transport particles from various scattering processes downstream into a set of concentric annuli; azimuthal-defocussing in the spectrometer produces a scattered flux distribution at the detector plane in the full range of the azimuth 30 m downstream of the target.

\begin{figure}
\begin{center}
\resizebox{0.95\textwidth}{!}{ \includegraphics{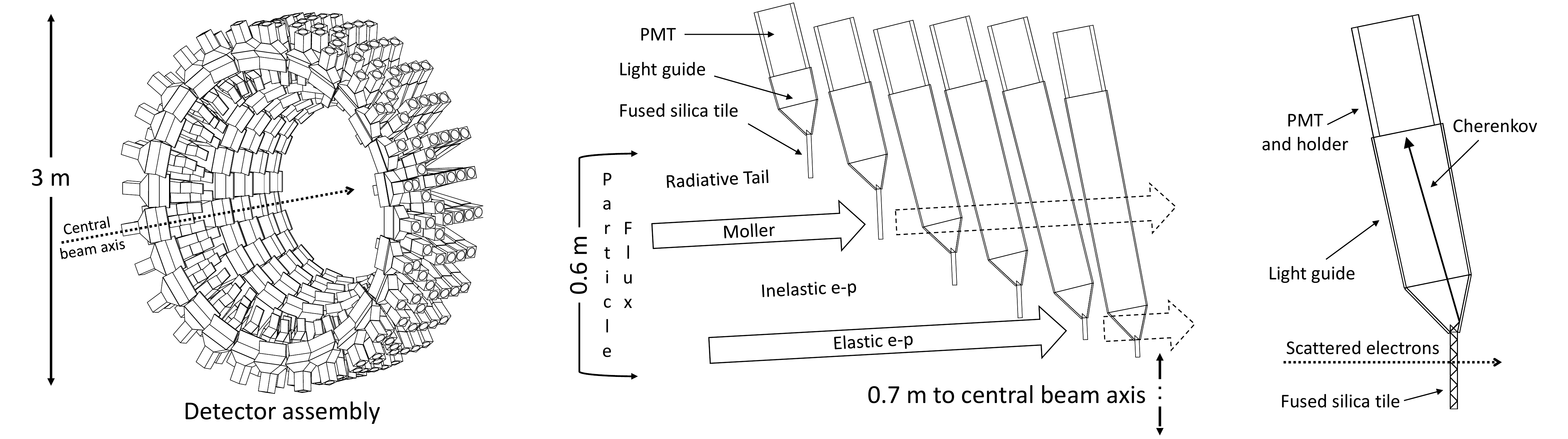} }
\caption{A schematic of the MOLLER experiment detector design. The full array is shown left. The different signals are focused into separate radial regions (center) and pass through the light guides of the following rings. A charged particle
passing through the fused silica tile while generating Cherenkov photons is 
illustrated in the right plot.}
\label{fig:MOLLER_detdiag}
\end{center}
\end{figure}

Figure \ref{fig:MOLLER_detdiag} (left) shows a schematic perspective view of the 
azimuthally and radially segmented detector array designed to intercept the 
electron annuli. Looking downstream along the beam axis, the
fused silica tiles form a series of concentric rings.
Each annulus will intersect electrons in different combinations scattering processes, such as M{\o}ller scattering, elastic electron-proton scattering, and inelastic electron-proton scattering. Coherent and incoherent photon radiation will additionally cause the migration of electrons into other detector regions.

Figure~\ref{fig:MOLLER_detdiag} (right) shows a diagram of how each tile is 
coupled to a long air light guide to channel tile-produced Cherenkov light to a PMT.
Figure~\ref{fig:MOLLER_detdiag} (center) shows a cross-sectional view of 
the tile, light guide, and PMT arrangement.
The rings are ordered such that larger radii are further upstream. Particles will therefore pass through their respective tile and subsequently all light guides belonging to smaller radius tiles.  This allows for the possibility that Cherenkov or scintillation light generated in the light guide volume will add to the signal from the tiles.  While the Cherenkov light yield, energy, and direction are readily calculable given an index of refraction, the detailed scintillation yield properties of gases are less well-known. Scintillation can also be a dominant contribution as it has an isotropic distribution and more difficult to geometrically suppress.  

A previous forward angle kaon experimental proposal known as KOPIO faced a similar problem and carried out a systematic analysis of scintillation yield in various gas mixtures~\cite{MORII2004399}. 
Their results suggest that the potential background in the MOLLER design with a PMT of a similar response
would be small, of order of 0.2\% in the main M{\o}ller rings, and could be as large
as 10-20\% in other rings. Such a background would dilute the parity-violating asymmetry and reduce the
effective analyzing power in various rings.

For a more accurate estimate of the background, a measurement was performed of the light yield from high energy electrons traversing various gases using a PMT with a similar wavelength response to those planned for use in the MOLLER experiment. 
Aside from air, the possibility of specialized gas mixtures was also explored.
In the following a description of the measurement apparatus and experimental configurations is provided.  An analysis of the collected data, a comparison to simulation, and our conclusions as applied to the MOLLER apparatus are also presented. 

\section{Scintillation Test}

\subsection{Detector Design}

\begin{figure}
\begin{center}
\resizebox{0.75\textwidth}{!}{ \includegraphics{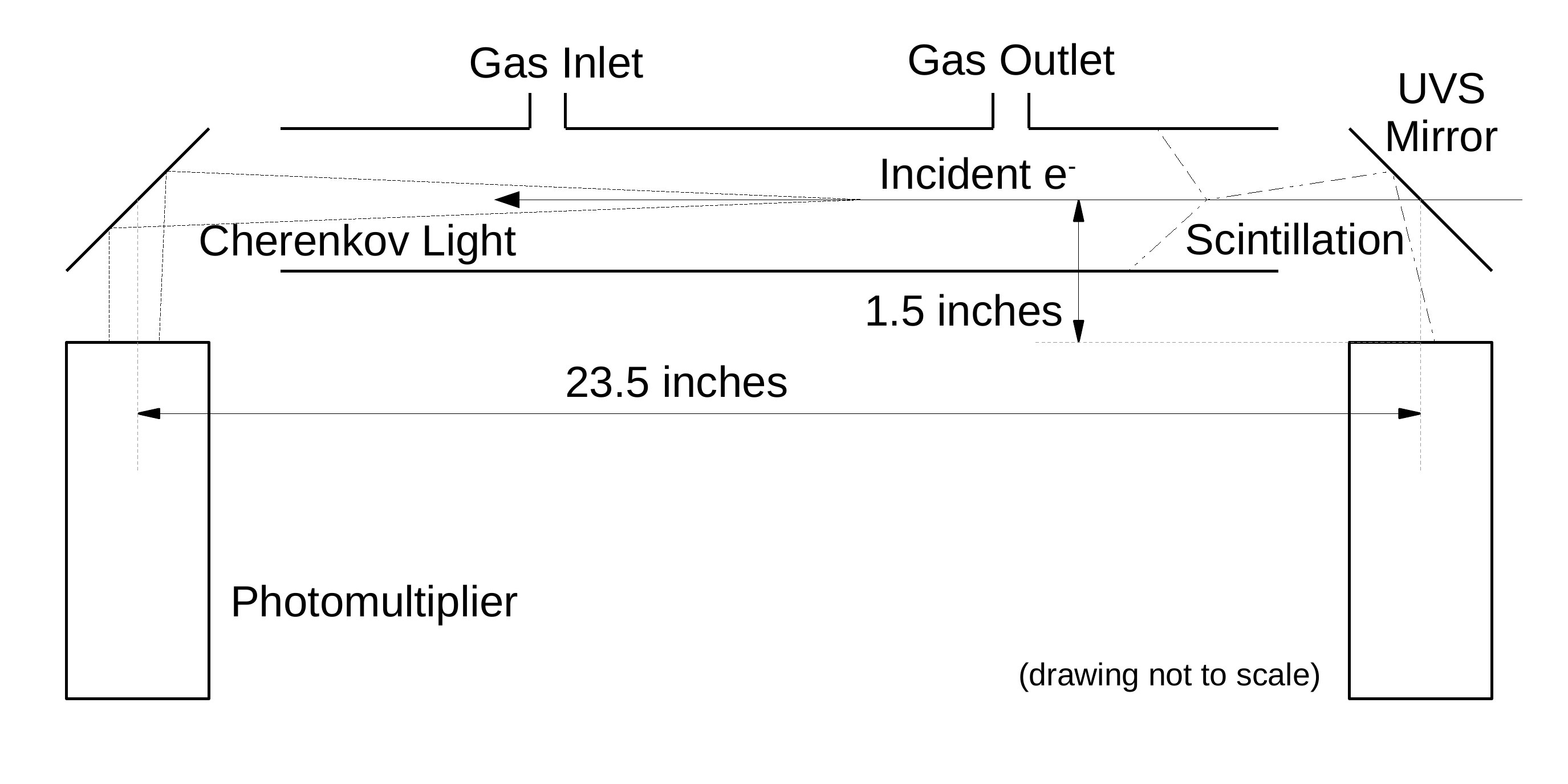} }
\caption{A schematic of the detector design.  An upstream PMT and UVS mirror at 45~degrees captures scintillation light and a matching pair downstream captures both Cherenkov and scintillation light from incident electrons.}
\label{fig:detdiag}
\end{center}
\end{figure}

\begin{figure}
\begin{center}
\resizebox{0.5\textwidth}{!}{ \includegraphics{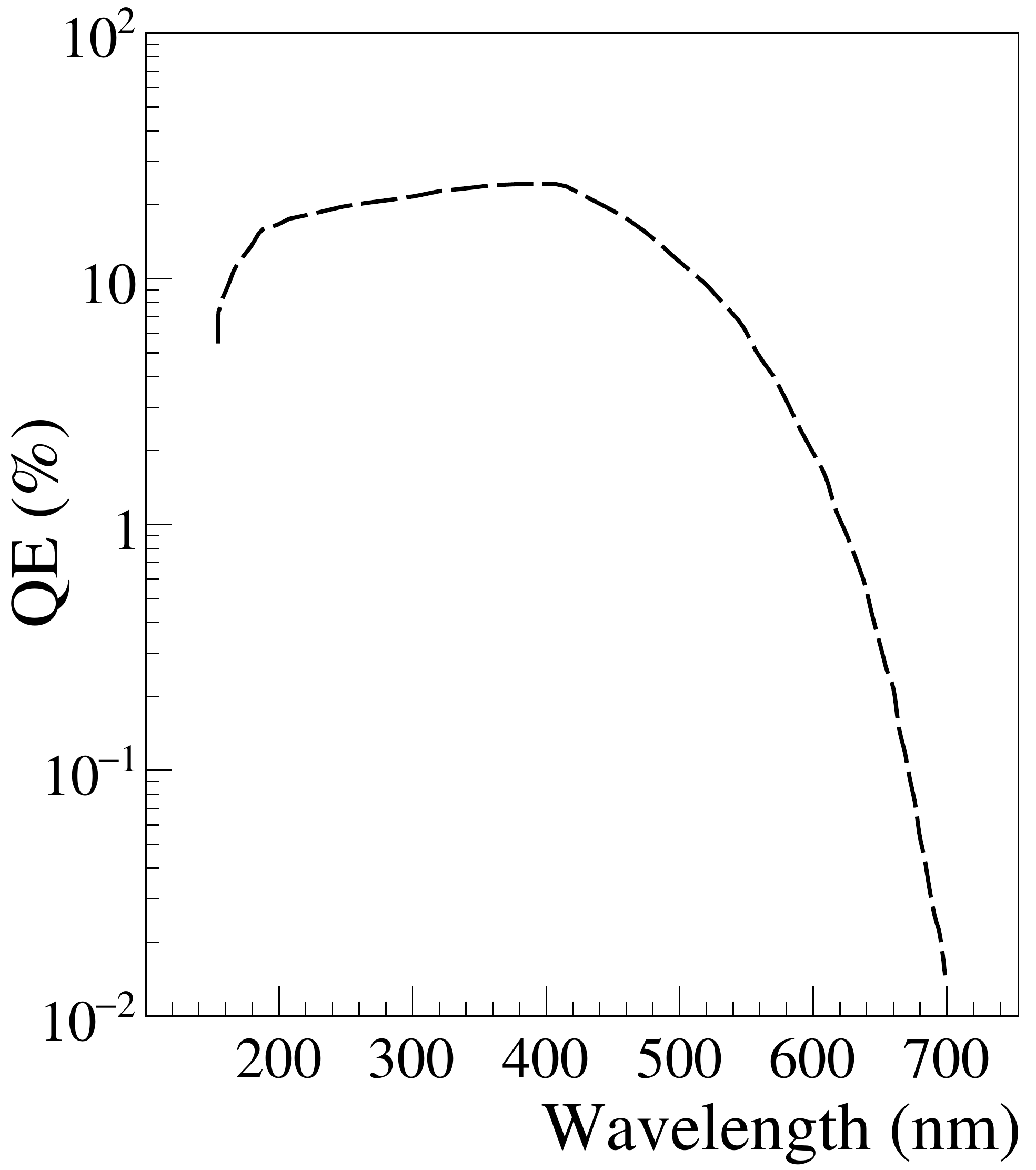} }
\caption{Quantum efficiency as a function of wavelength
for the Hamamatsu~H3177 PMTs.}
\label{fig:PMT}
\end{center}
\end{figure}

\begin{figure}
\begin{center}
\resizebox{0.6\textwidth}{!}{ \includegraphics{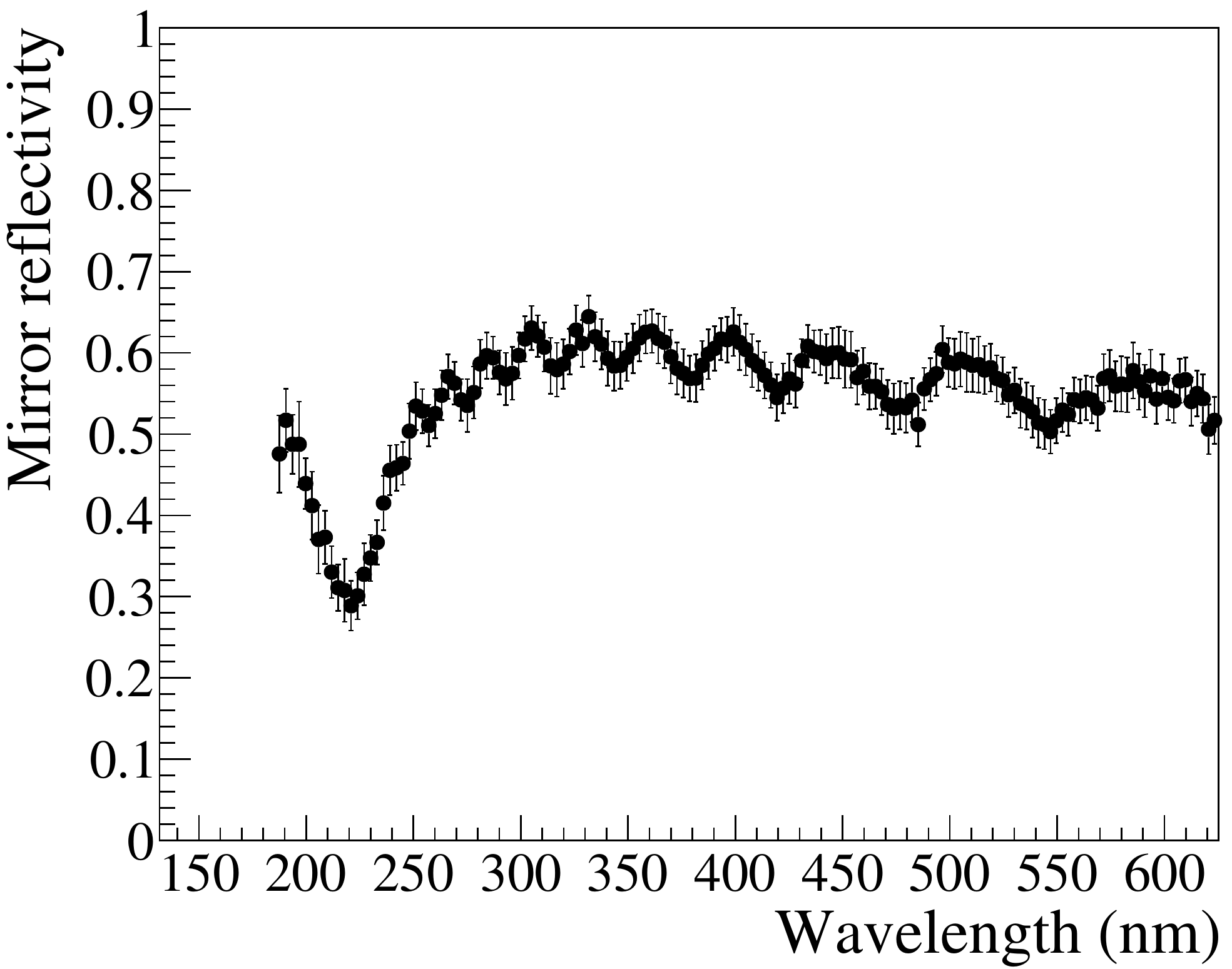} }
\caption{The measured reflectivity of the mirror
as a function of wavelength for incident angle 
of 45 degrees. The error bar represents the combined statistical and systematic uncertainties.}
\label{fig:mirror}
\end{center}
\end{figure}

To measure the Cherenkov and scintillation yields, a detector was constructed which consists of a tube filled with gas whose axis is collinear with an incident electron beam, as shown in Figure~\ref{fig:detdiag}.  Anolux UVS mirrors angled 45~degrees to the axis are placed on each end which serve to reflect generated light into a PMT which lies 90~degrees to the axis. 
The test was carried out using two Hamamatsu~H3177 PMTs. The quantum
efficiency as a function of wavelength provided by the manufacturer is shown in Figure~\ref{fig:PMT}. 
The reflectivity of the mirror, as shown in Figure~\ref{fig:mirror}, was measured at an incident angle of 45~degrees for photons over a range of optical and ultraviolet wavelengths.
The upstream PMT will predominantly detect isotropically-produced scintillation light and the downstream PMT will detect both scintillation light and Cherenkov light. In practice, reflections from the downstream PMT will propagate some Cherenkov signal to the upstream PMT and is accounted for by collecting data with the downstream PMT and mirror removed and the aperture sealed with a non-reflective surface. The calculable Cherenkov signal serves as a known reference quantity.

The inner diameter of the tube was about 2~in, which is slightly larger than the chosen PMT radius. The amount of collected scintillation light is a combination of length of the detector and the inverse radius-squared distance from where the light is produced.  An interaction length of about 23.5~in was chosen so that the largest Cherenkov ring (produced by the most upstream electrons) would match the size of the PMT photo-cathode.  

The integrated Cherenkov production per unit energy per electron can be approximated by
\begin{equation}
    \frac{d^2N}{dEdx} \approx 370 \sin^{2} \theta_c ~ \mathrm{eV}^{-1} \mathrm{cm}^{-1}
\end{equation}
where $\theta_c$ is the Cherenkov angle, and $\cos \theta_c = c/(n v)$ where $n$ is the index of refraction, $c$ is the speed of light, and $v$ is the particle velocity. For pure nitrogen at atmospheric pressure, an electron energy of at least $100~\mathrm{MeV}$, and a $3.5~\mathrm{eV}$ photon energy acceptance width, will produce about 40 photons.

The scintillation yield assuming isotropic uniform production, a beam coincident with the detector axis, and perfect detection efficiency is approximately given by
\begin{equation}
    N = \lambda \frac{a^2}{16 d} \tan^{-1} \left( \frac{L + d}{d} \right)
\end{equation}
with a yield per unit length $\lambda$, a PMT with aperture $a$ of distance $d$ from the mirror, and a detector of length $L$. The total optical length from PMT faces for our configuration is 26.5~in and collects more than 95\% of the scintillation light as an equivalent detector of infinite length.

The main body of the detector was constructed using black PVC plastic piping and fittings with a minimum internal radius of 2.07~in and nominal wall thickness of 0.154~in. The interior of each tube was roughed with sand paper to minimize internal reflection. A sleeve for the PMT was constructed using the same piping with the internal radius machined to match the PMT outer radius.  Silicone O-rings of thickness 3/32~in  were glued into place where the PMT face would make direct contact with the piping. Mirrors were constructed from Anolux UVS and glued into place on an elbow fitting cut at $45$~degrees.

The system has two $1/4$~in inlets to allow gas flow controlled by manual quarter-turn plug valves.  Black opaque silicone tubing was used to help ensure light-tightness.  The system was not designed as a pressure vessel or to have gas tightness.  Internal positive gas pressure slightly above atmospheric pressure was ensured by an oil bubbler on the outlet line. An extruded aluminum frame was constructed to hold the detector in fixed position as well as to mount it to a fixed plate.

\subsection{Test Beam Setup}
\label{section_beam_test}

\begin{figure}
\begin{center}
\resizebox{0.95\textwidth}{!}{ \includegraphics{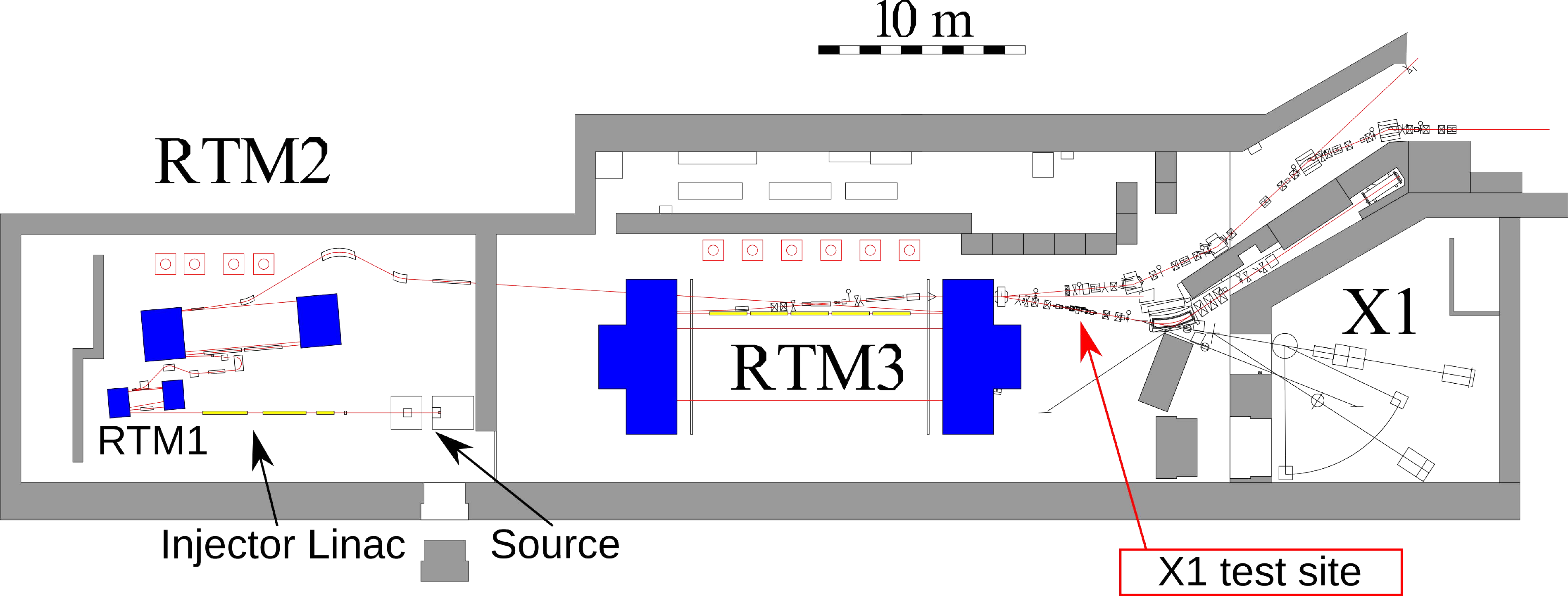} }
\caption{(Color online) The floor-plan of the first accelerator stages of the MAMI facility in Mainz, Germany with the test site (labeled X1) used for the scintillation test studies. }
\label{fig:RTM3floorplan}
\end{center}
\end{figure}

\begin{figure}
\begin{center}
\resizebox{0.6\textwidth}{!}{ \includegraphics{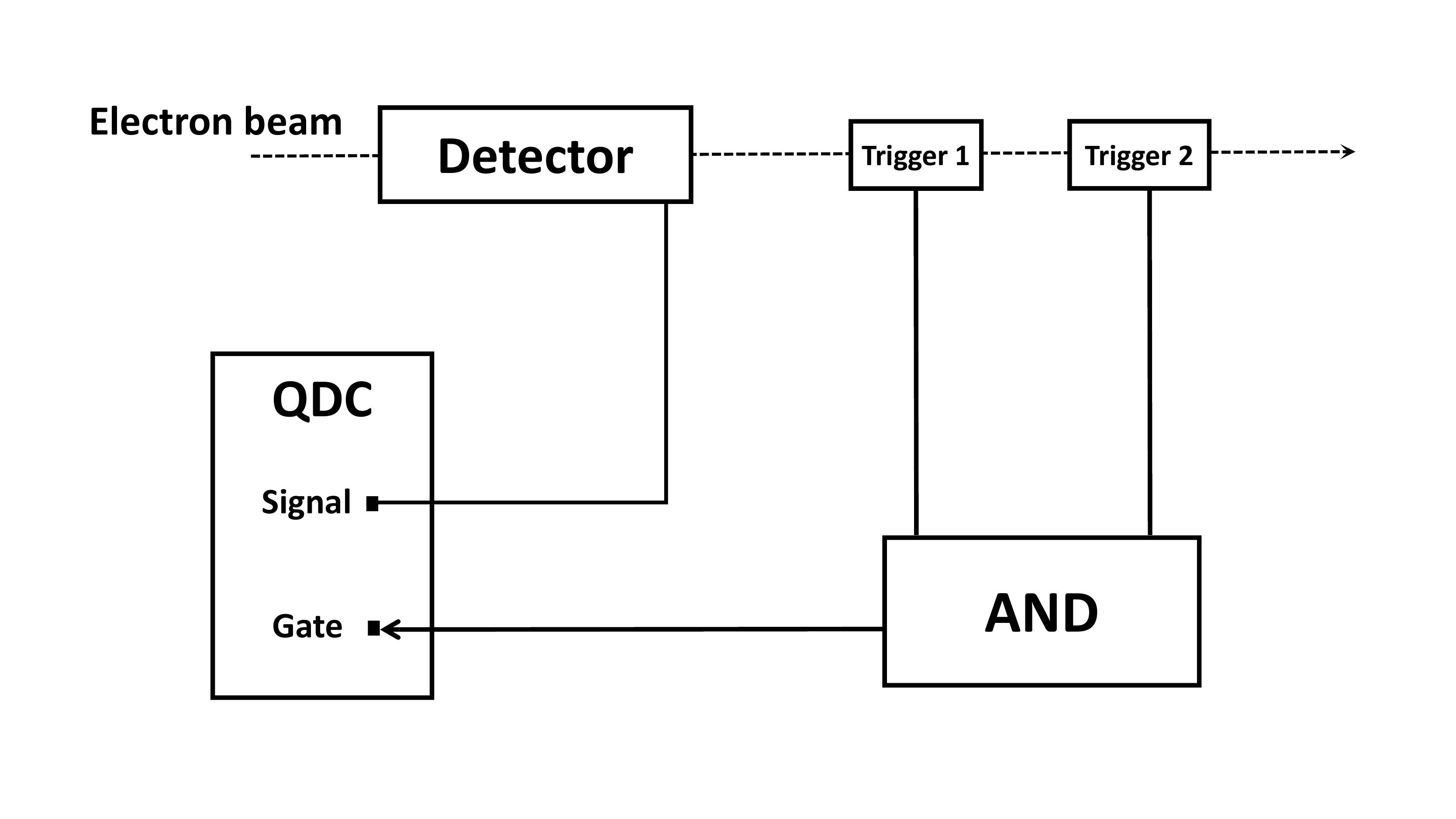} }
\caption{A diagram of the trigger formation for the light measurement. }
\label{fig:Triggerschaltung}
\end{center}
\end{figure}

The measurements were performed at the MAMI electron accelerator facility in Mainz, Germany at the X1 test site, which is situated after the third Racetrack Microtron (RTM3). A schematic view of the test area is shown in Figure~\ref{fig:RTM3floorplan}. After 90 circulations in the RTM3, the electrons reach an energy of 855~MeV. The detector was mounted on an electronically-controlled motion table under the beam line.  Two pieces of plastic scintillator connected to PMTs were mounted downstream of the detector which formed a trigger under a coincidence of only these two signals, as shown in Figure~\ref{fig:Triggerschaltung}.  A CAEN V965 QDC records the integrated charge from the main detector PMTs. 

The detector was filled with several different gases, including air, N$_2$, CO$_2$, and Ar with a dedicated gas circulation system. 
Prior to the test, gain calibrations for the two main detector PMTs were performed.  The
nominal PMT operating high voltages were then determined by requiring that the single 
photo-electron (P.E.) peak can be distinguished 
from the pedestal peak in a QDC with 200 fC/channel resolution.
For each gas test, the PMTs were operated with different high voltages, 2600V, 2700V, and 2800V. 
In addition to the two-PMT configuration, so-called ``block runs'' were performed with the downstream mirror and PMT removed and the opening covered with black paper.  These runs allowed for a determination of the amount of reflected Cherenkov light from the downstream to upstream PMT.  The ``block runs'' were only performed with the detector filled with air.  For each run during the beam test, at least one million events were taken.

\section{Data Analysis and Simulations}

\subsection{Test Beam Results}
\begin{figure}
\begin{center}
\resizebox{0.75\textwidth}{!}{ \includegraphics{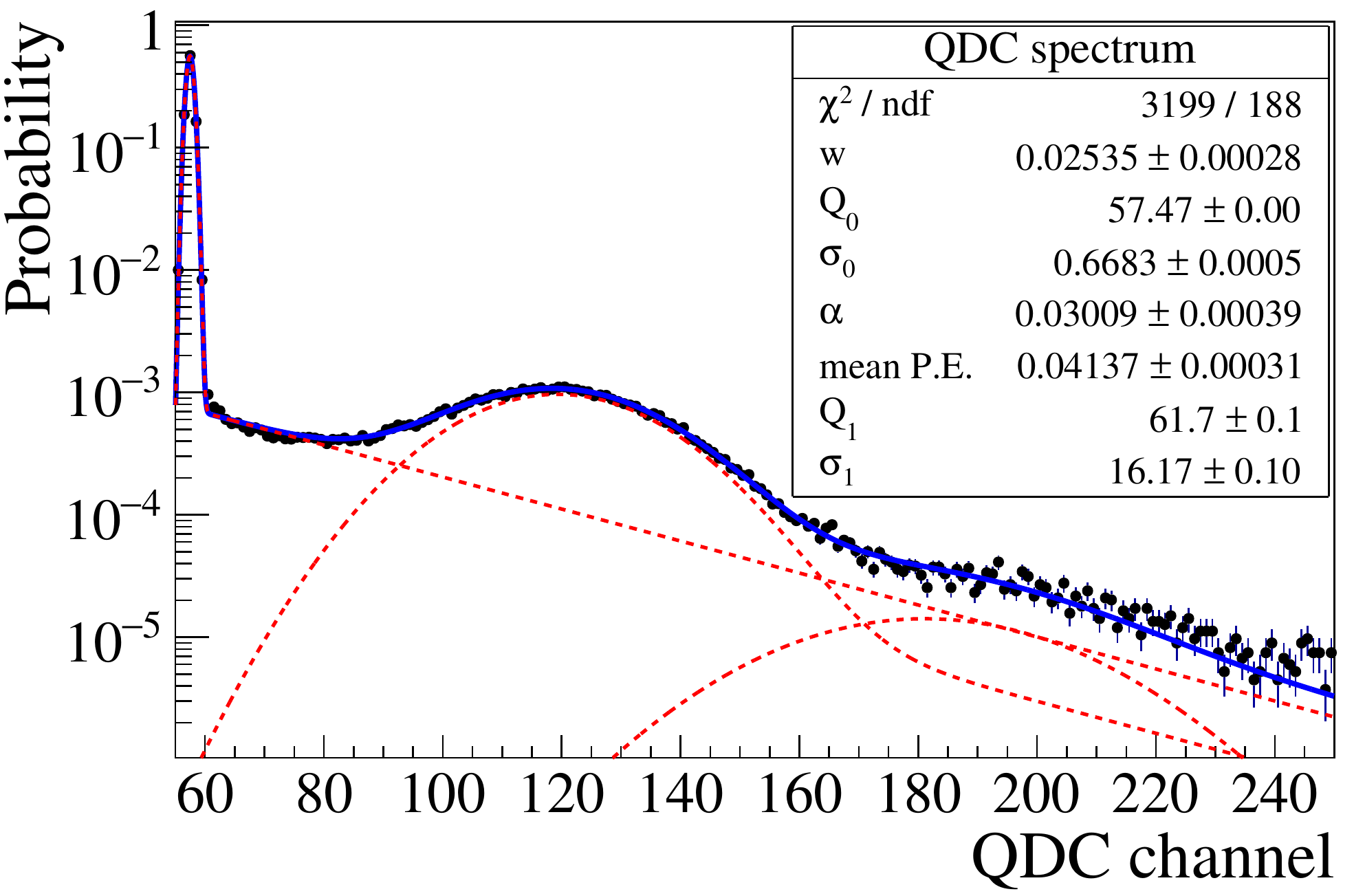} }
\caption{(Color online) A sample fit to a QDC spectrum for air from the upstream PMT. The black points are data from the test. The blue line is the total fit to the data while the red dash peaks show the contributions from individual photo-electrons in the fit. Q$_0$ is the central position for the pedestal, $\sigma_{0}$ is the width of the pedestal peak, Q$_{1}$ is the difference between the central position of the single photo-electron peak and pedestal, and $\sigma_{1}$ is the width of the single photo-electron peak. $w$ is the probability that a measured signal is accompanied by discrete background processes from the PMT, such as thermo-emission or noise initiated by the measured light. $\alpha$ is the coefficient of the exponential decrease of the discrete background processes \cite{Bellamy:1994bv}. }
\label{fig:scint_sample}
\end{center}
\end{figure}

\begin{figure}
\begin{center}
\resizebox{0.75\textwidth}{!}{ \includegraphics{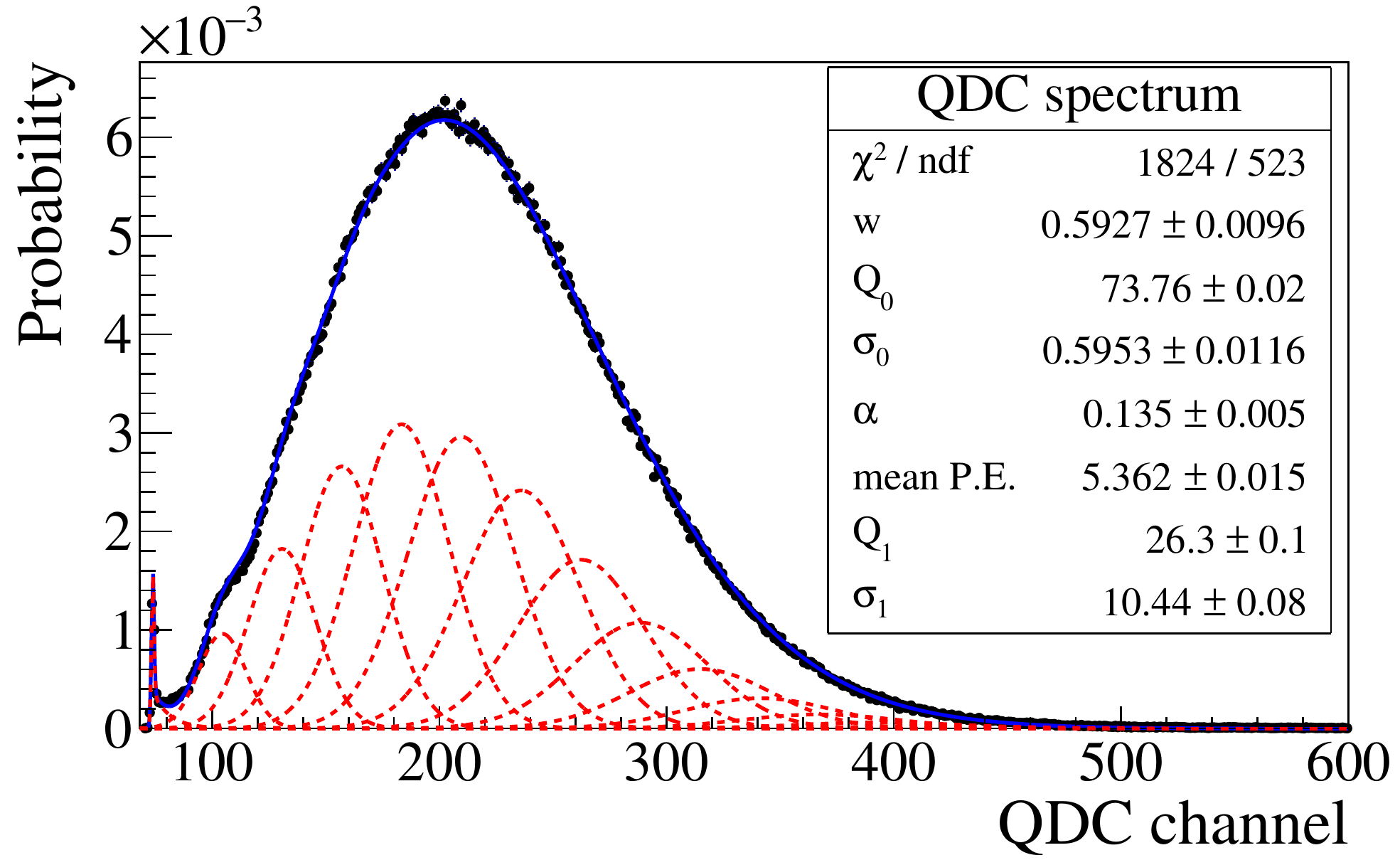} }
\caption{(Color online) A sample fit to a QDC spectrum for air from the downstream PMT. The black points are data from the test. The blue line is the total fit to the data while the red dash peaks show the contributions from individual photo-electrons.  The meaning of the fit parameters shown in the inset are the same as in Figure~\ref{fig:scint_sample}.}
\label{fig:cere_sample}
\end{center}
\end{figure}

The observed signals are significantly less than one photo-electron on average for the upstream PMT and a few photo-electrons for the downstream PMT. A fit which was developed for the absolute gain calibration of PMTs~\cite{Bellamy:1994bv} using Poisson statistics was employed to analyze the data. The QDC spectrum was fit directly to obtain the expected mean number of P.E.'s which accounts for details of the PMT response. Figure~\ref{fig:scint_sample} is a sample fit to a QDC spectrum collected by the upstream PMT and shows contributions from up to two photo-electrons with a mean value of less than 0.1 P.E. The exponential tail is attributed to the PMT response due to thermo-emission and noise initiated by the measured light~\cite{Bellamy:1994bv}. 
Figure~\ref{fig:cere_sample} shows an example fit to the Cherenkov signal on the downstream PMT.

\begin{table}
\begin{center}
\begin{tabular}{|c|c|c|c|c|}\hline
    Gas type   &  Averaged mean P.E.  & Standard deviation & Averaged mean P.E. & Standard deviation   \\      
    & Upstream PMT &   Upstream PMT & Downstream PMT &   Downstream PMT   \\ \hline 
Air               &     0.0409    &   0.0007 (1.7\%)      & 5.4     &  0.3 (4.7\%)     \\ \hline
N$_2$             &     0.070     &   0.002 (2.2\%)       & 5.8     &  0.1 (1.5\%)   \\ \hline
CO$_2$            &     0.0440    &   0.0007 (1.5\%)      & 7.4     &  0.4 (5.9\%)  \\ \hline
Ar                &     0.181     &   0.002 (1\%)         & 5.7     &  0.2 (3.8\%)    \\ \hline
\end{tabular}
\caption{The tabulated numbers of the averaged number of mean P.E.'s for different gases. The numbers are directly obtained from the signals collected from the upstream or downstream PMT as noted. The standard deviation of means for the collection of similar runs is also shown. The number in the parenthesis is the ratio of the standard deviation to the mean value in percent. }
\label{tab:PMT_signals}
\end{center}
\end{table}

\begin{figure}
\begin{center}
\resizebox{0.75\textwidth}{!}{ \includegraphics{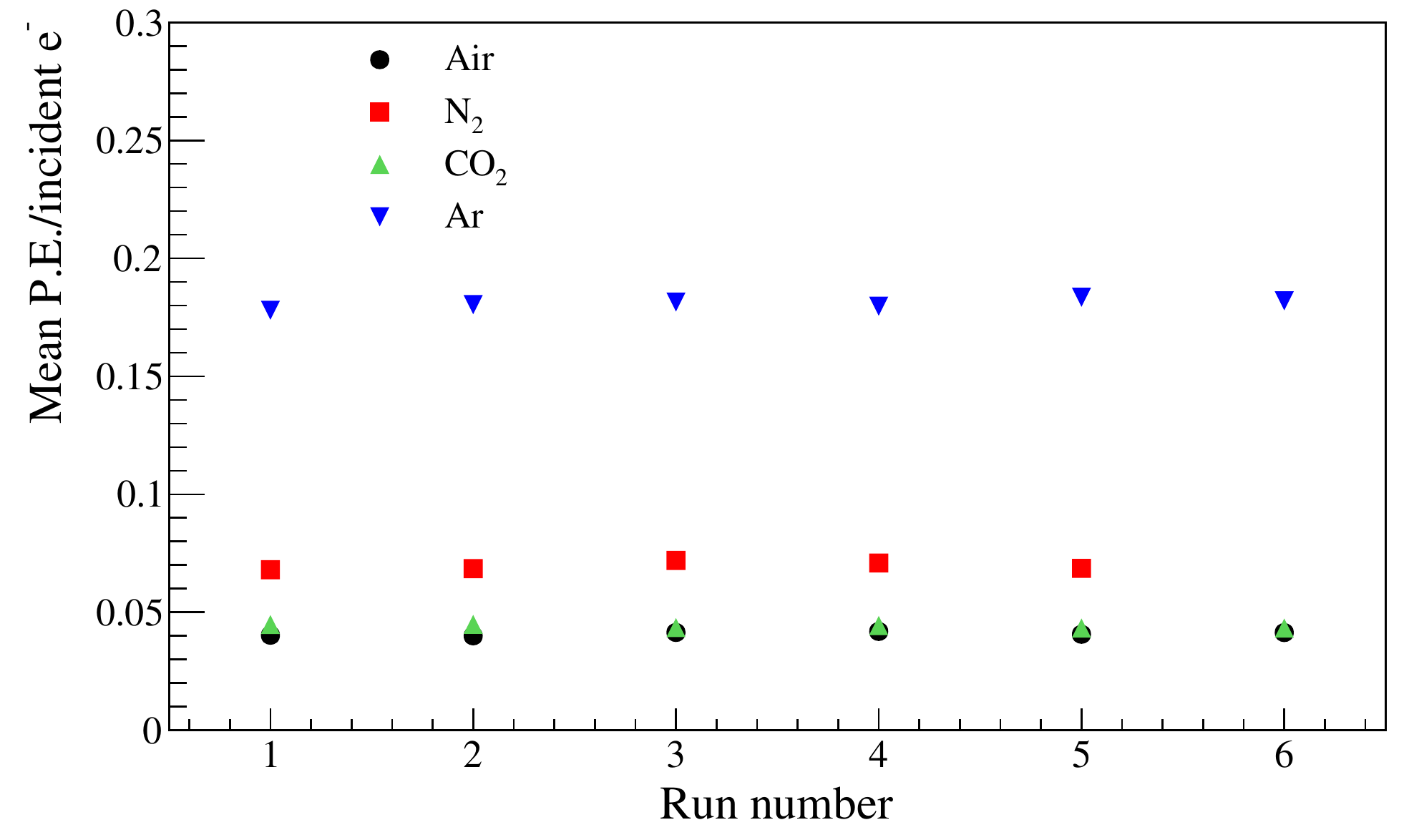} }
\caption{(Color online) The mean photo-electrons on the upstream PMT for different runs. These
numbers contain both the direct scintillation signal and reflected Cherenkov light from the downstream mirror and PMT.}
\label{fig:upstream_PMT}
\end{center}
\end{figure}

\begin{figure}
\begin{center}
\resizebox{0.75\textwidth}{!}{ \includegraphics{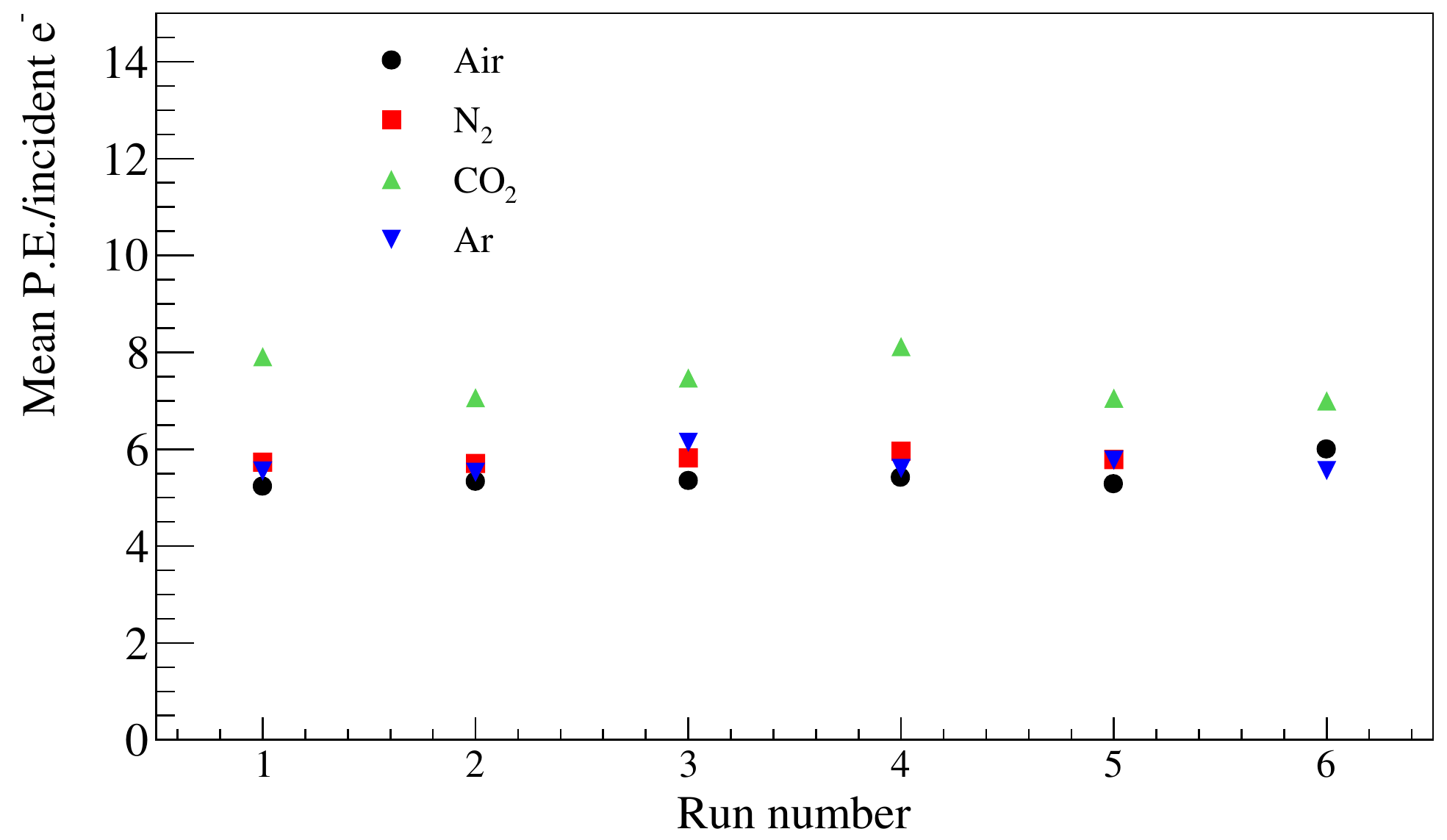} }
\caption{(Color online) The mean photo-electrons on the downstream PMT for different runs. These
numbers are predominantly measurements of the Cherenkov signal.}
\label{fig:downstream_PMT}
\end{center}
\end{figure}

Several runs of about 1~million events each for each of the four different gases were taken. The fitted mean numbers of P.E.'s for the upstream and downstream PMT are shown in Figures~\ref{fig:upstream_PMT} and~\ref{fig:downstream_PMT} respectively and are summarized in Table \ref{tab:PMT_signals}.  The standard deviation of the extracted mean number of P.E.'s between runs was typically found to be better than 0.002 P.E. for the scintillation signal ($\sim 2$\%), a remarkable achievement demonstrating the stability and robustness of
the measurement technique. For the larger Cherenkov signal, the standard deviation was 0.4 P.E. or better ($\sim 5$\%). 

The detected signal in the upstream PMT contains contributions from both the scintillation and Cherenkov light reflected by the downstream mirror and PMT.  This is particularly evident as the CO$_2$ scintillation signal is expected to be negligible.  A measure of the reflected light efficiency was obtained by the ``block run'' described above.  For air, the mean number of upstream P.E.'s obtained with two PMTs present is 0.041, while the mean number of upstream P.E.'s is 
0.0086 $\pm$ 0.0001 (statistical uncertainty) with the black paper in place, demonstrating that 
the dominant contribution in the upstream PMT for all gases but Argon is from reflected Cherenkov light. Thus, a subtraction
is crucial to obtain a meaningful measure of the scintillation response. This is also confirmed in the Geant4 simulation, as will be discussed in Section~\ref{section_simulation}.

\subsection{Comparison to Simulations}
\label{section_simulation}

The experimental and detector configuration was implemented in a Geant4~\cite{AGOSTINELLI2003250} simulation and 
the output was compared to the data. This serves to validate our understanding of the involved processes
as well as geometric dependencies and provides a benchmarked simulation that can be used
to study the case of the MOLLER detector geometry.

\begin{figure}
\begin{center}
\resizebox{0.75\textwidth}{!}{ \includegraphics{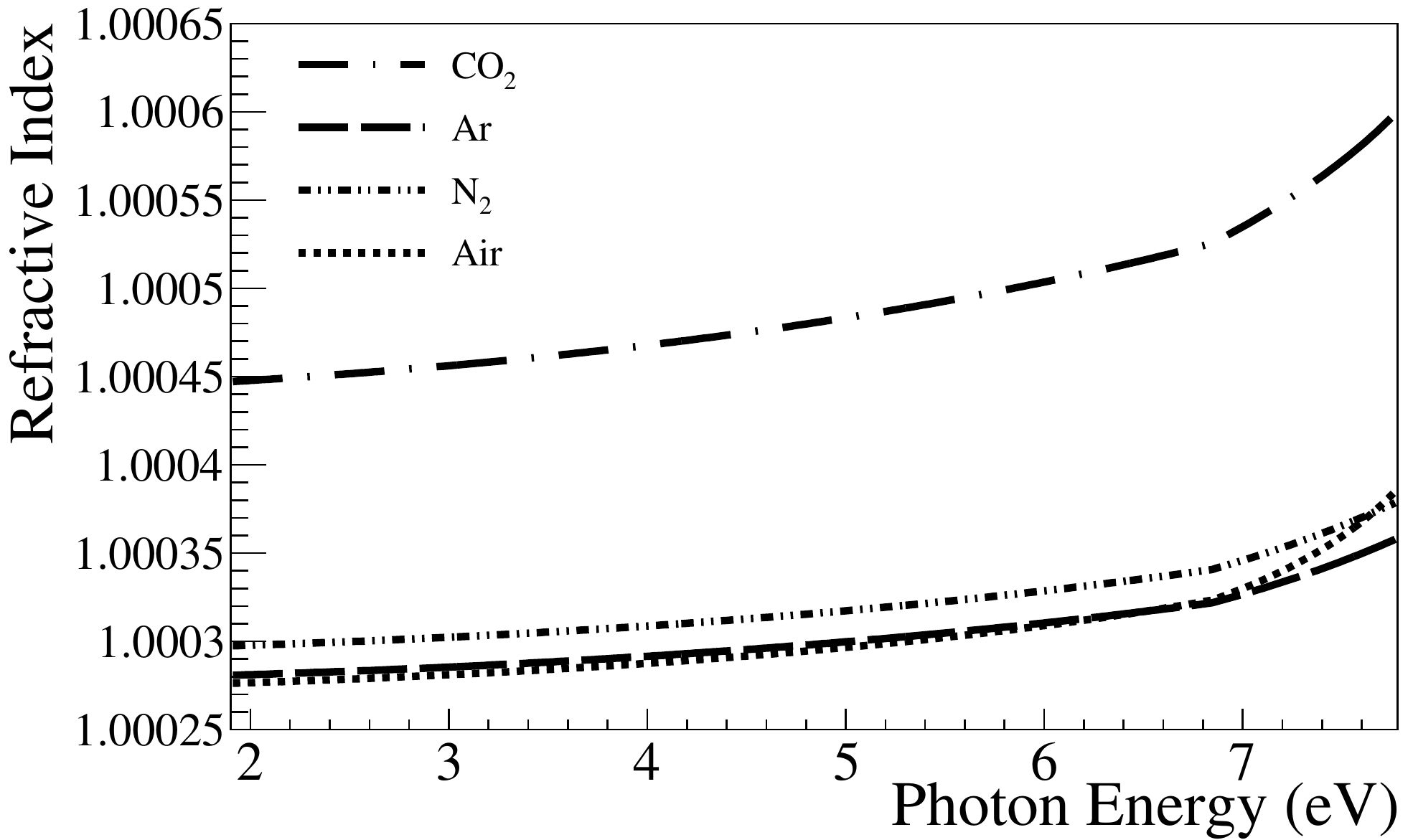} }
\caption{The refractive index used in the simulation for different types of gas as a function of photon energy~\cite{rii}.}
\label{fig:refractive}
\end{center}
\end{figure}

The simulation covers a photon wavelength region from 180 to 600 $nm$. The quantum efficiency of the PMTs and the mirror reflectivity with 45$^\circ$ incidence, both as functions of photon energy, are included. For Cherenkov light emission, the refractive index used in the simulation is shown in Figure \ref{fig:refractive}. For gas scintillation, the light yield is implemented for different gas types for which the numbers are tabulated in Table \ref{tab:scint_light_yield}. 
The numbers for air, N$_2$ and CO$_2$ are available explicitly in 
the reference \cite{MORII2004399}, while the number for Ar gas is 
an estimation based on the information in references \cite{MORII2004399}
and \cite{GERNHAUSER1996300}.
Geant4 allows one to turn on scintillation and Cherenkov processes separately or together. With both processes on, one can directly compare the simulation results with the beam test results. After the comparison, one can study individual processes.

\begin{table}
\begin{center}
\begin{tabular}{|c|c|c|c|c|}\hline
    Gas type:            &    Air   &    N$_2$    &   CO$_2$    &   Ar       \\      \hline
Yield(photons/MeV):       &    25    &    140      &    5        &   510     \\ \hline
\end{tabular}
\caption{The assumed scintillation light yield for each type of gas \cite{MORII2004399,GERNHAUSER1996300}.}
\label{tab:scint_light_yield}
\end{center}
\end{table}

The signals from both of the two PMTs were simulated and the mean values for the number of P.E.'s obtained for different types of gas are summarized in Table \ref{tab:sim_scin_cere}. The overall agreement is 
satisfactory for the purposes of controlling the relative uncertainty of the light guide background over signal at the level of 10\%.

\begin{table}
\begin{center}
\begin{tabular}{|c|c|c|c|c|}\hline
    Gas type      &     Simulated mean P.E. & Data mean P.E.  & Simulated mean P.E.  & Data mean P.E.  \\  
                   &     Upstream PMT       &   Upstream PMT     &    Downstream PMT   & Downstream PMT  \\ \hline
Air               &     0.039       &  0.041      & 6.0   &   5.4   \\ \hline
N$_2$             &     0.069       &  0.07       & 6.3   &   5.8   \\ \hline
CO$_2$            &     0.021       &  0.044      & 9.6   &   7.4   \\ \hline
Ar                &     0.17        &  0.18       & 6.0   &   5.7  \\ \hline
\end{tabular}
\caption{The mean number of P.E.'s for different gases in each PMT from the simulation. The results from the measurement as presented in Table \ref{tab:PMT_signals}, are also shown for comparison. }
\label{tab:sim_scin_cere}
\end{center}
\end{table}

In addition, the pure scintillation light yield was also studied using the ``block runs'' to constrain the reflected Cherenkov light. The reflection efficiency factor obtained is applied to all configurations, assuming that this factor is independent of medium. In order to simulate the pure scintillation yield in the upstream PMT, the Cherenkov process was turned off. Table~\ref{tab:pure_scint} shows the comparison of pure scintillation signal between the measurement and simulations.
The data obtained during the
beam test with air agree with the simulation at the level of 0.003 photo-electron; we estimate the sensitivity of this method, given the statistical precision and systematic control to be about 0.002 photo-electron. 

It is worth noting that the background subtraction yields a value close to zero for CO$_2$, as should be the case.
 In addition to pure gas, gas mixtures such as Ar(95\%)+CO$_2$(5\%) or Ar(90\%)+CO$_2$(10\%) were also tested.
We observed that the signal in the two Argon,CO$_2$ mixtures could be completely explained as reflected Cherenkov light, i.e., the scintillation response of argon with 5 to 10\% CO$_2$ was consistent with zero within the measurement precision and systematic control. The N$_2$ scintillation signal is clearly larger than air by a factor of 3 to 6, showing the importance of quenching of the N$_2$ signal by other gases in the air. 

\begin{table}
\begin{center}
\begin{tabular}{|c|c|c|}\hline
    Gas type      &     Beam test results  & Simulation results          \\      \hline
Air               &     0.0086             & 0.005           \\ \hline
N$_2$             &     0.035              & 0.033         \\ \hline
CO$_2$            &     $\sim$0 ($<$0.001)              & 0.001      \\ \hline
Ar                &     0.15               & 0.14     \\ \hline
\end{tabular}
\caption{A comparison of pure scintillation signal between measured data and simulations.}
\label{tab:pure_scint}
\end{center}
\end{table}

\section{Estimation of MOLLER Detector Scintillation Background}

After benchmarking the Geant4 simulation of the scintillation test tube, the geometry of a 
real detector which is proposed for use in the MOLLER experiment was constructed in the 
simulation using the same parameters. 
The detector is constructed out of a fused silica Cherenkov radiator, a mirror funnel and 
a long light guide with a PMT in the end. 
As illustrated in Figure \ref{fig:MOLLER_detdiag} (right), the dominant signal in the PMT will be
generated by high energy electrons passing through the fused silica tile and the Cherenkov 
light will be released in a cone of half-angle of about $45^{\circ}$. Through repeated total 
internal reflections, a significant fraction of the light will reach the $45^{\circ}$ cut at 
one end of the radiator, where the light will enter the light guide. Per the design of the 
detector, the light will be reflected to the PMT by the mirror attached to the radiator 
via only one bounce.  

\begin{figure}
\begin{center}
\resizebox{0.75\textwidth}{!}{ \includegraphics{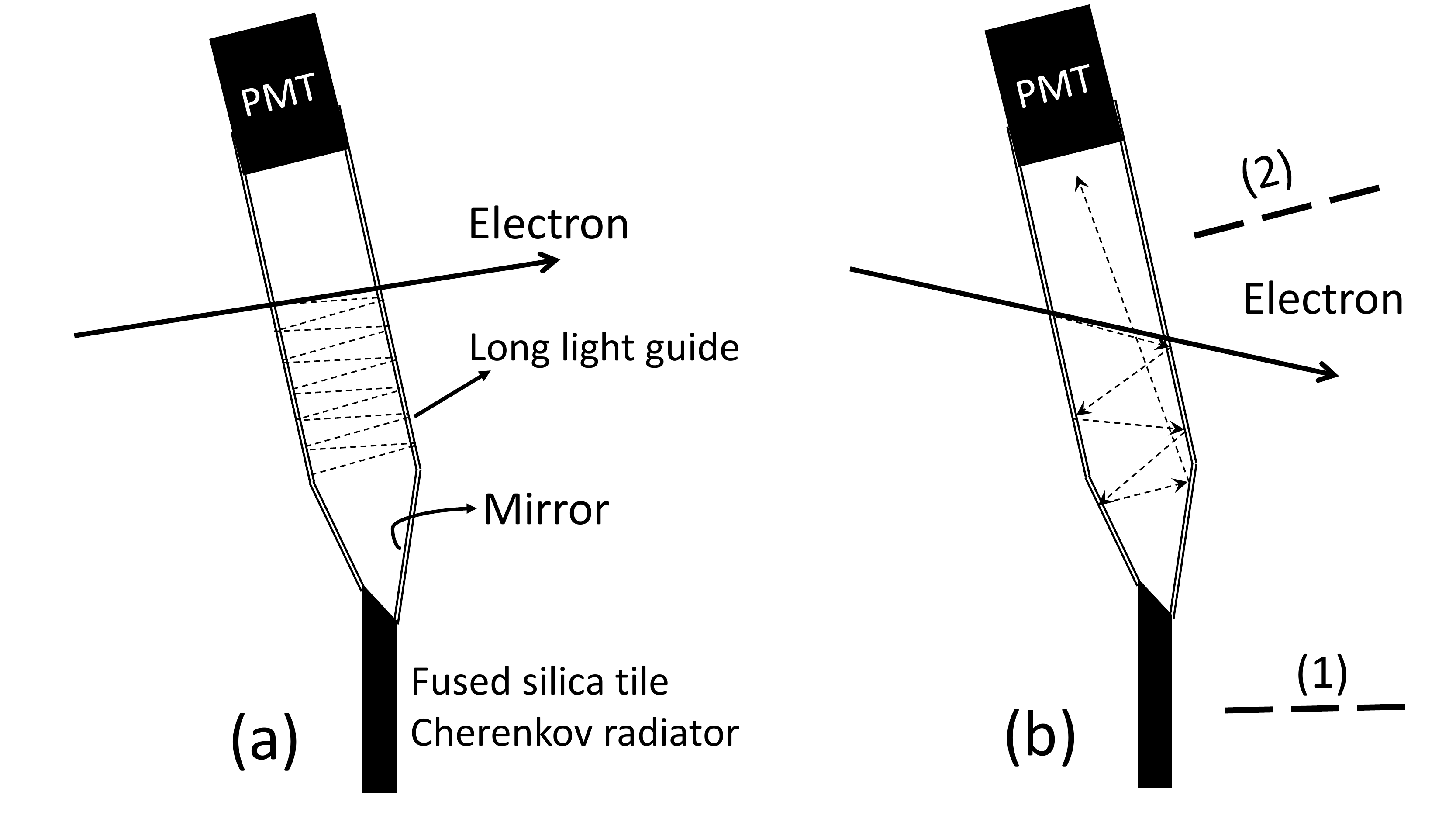} }
\caption{An illustration of the detector module proposed in the MOLLER experiment and a depiction of the background
Cherenkov light generated by charged particles going through a neighboring light guide.
Line (1) in figure (b) is normal to the fused silica tile while line (2) is normal to the 
surface of the long light guide. The angle between (1) and (2) is 11.5 degrees.}
\label{fig:real_detector}
\end{center}
\end{figure}

\begin{figure}
\begin{center}
\resizebox{0.75\textwidth}{!}{ \includegraphics{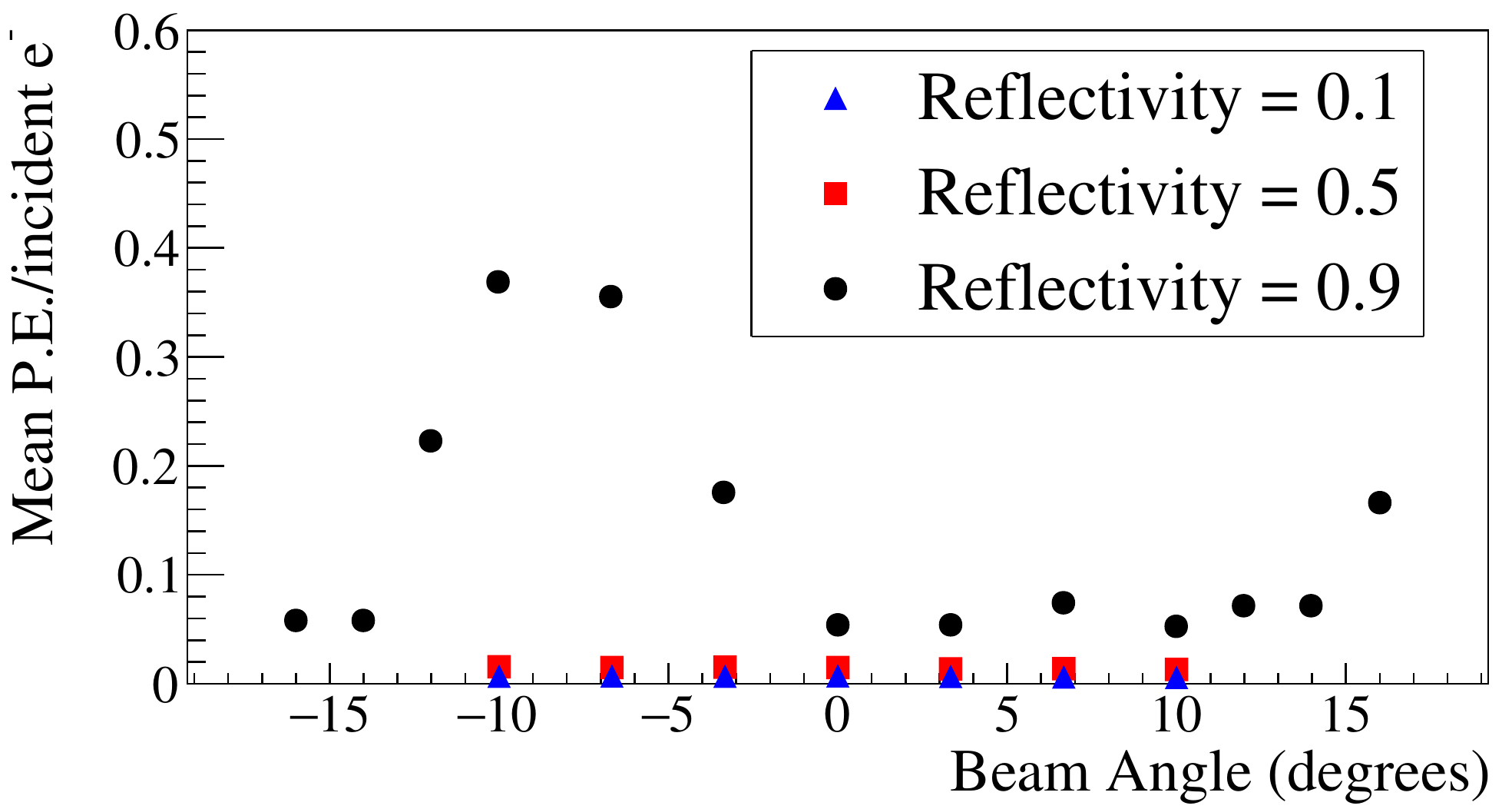} }
\caption{(Color online) The Cherenkov and Scintillation light yield for beam angle scan with an air-filled light guide . The incident electrons are shot along the long light guide.  The angle is relative to the surface of radiator tile with an angle of zero being parallel to line (1) and positive angle in the direction of line (2), both shown in Figure \ref{fig:real_detector} (b).
}
\label{fig:detector_sim_both}
\end{center}
\end{figure}

To evaluate the main signal, the light yield from individual electrons passing through the tile 
was simulated.  For the background, the electron beam is shot directly on the long light guide 
filled with air. As illustrated in Figure \ref{fig:real_detector}, a scan over incident angles was performed 
with both Cherenkov and scintillation 
processes turned on. Figure \ref{fig:detector_sim_both} shows the mean number of P.E.'s as a 
function of beam angle relative to the surface of the radiator tile.  An angle of zero 
corresponds to the beam normal to the surface of the tile. Positive angle corresponds to the 
beam pointing towards the PMT. The study was performed for different reflectivities of the 
light guide. As shown in Figure \ref{fig:detector_sim_both}, the signal is 
strongly suppressed with low reflectivity. The rising effect near $-10$~degrees with a 
reflectivity of 0.9 is due to the fact that the air-produced Cherenkov light is reflected 
towards the PMT with only a few bounces, as shown in Figure \ref{fig:real_detector} (b) 
compared to Figure \ref{fig:real_detector} (a), by the mirror attached to the radiator tile. 
At other incident angles, the light yield will drop due to the attenuation of many bounces 
inside the light guide before photons reach the PMT surface.  The light yield also increases 
in the region where the beam angle is larger than +10$^{\circ}$, since the beam is 
pointing more towards the PMT.

\begin{figure}
\begin{center}
\resizebox{0.75\textwidth}{!}{ \includegraphics{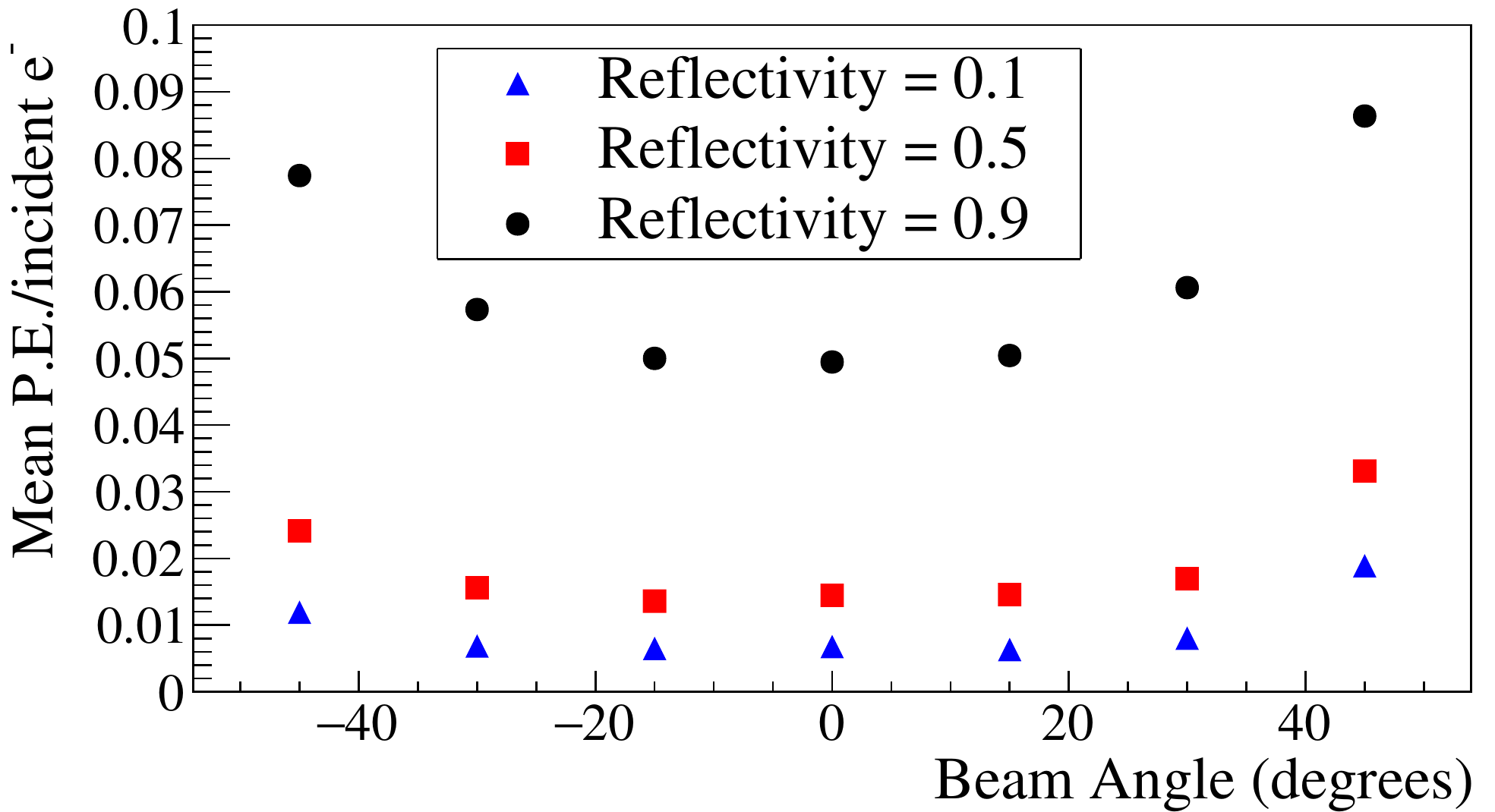} }
\caption{(Color online) Scintillation yield for beam angle scan. Air is filling the light guide. The electron beam is shot on the light guide directly.  The angle is relative to the surface of the light guide with an angle of zero parallel to line (2) in Figure \ref{fig:real_detector} (b) and negative angle in the direction of (1).}
\label{fig:detector_scint}
\end{center}
\end{figure}

A scan over incident angles was also performed with only scintillation. Figure \ref{fig:detector_scint} shows the mean number of P.E.'s as a function of beam angle. Zero angle in this case means that the electron is normal to the surface of the light guide.  This corresponds to the minimum scintillation light yield due to the length of the radiator traversed.  The scintillation yield is much smaller than the multi-bounce Cherenkov light. 
 
In the MOLLER design shown in Figure \ref{fig:MOLLER_detdiag}, the detectors with tiles at a smaller radius will receive a larger flux in their longer light guides.
Based on our benchmark simulations, we have carried out a cursory study
of the light guide background for each of the rings. 
We note that the Cherenkov background can be strongly suppressed by blackening a small portion of the 
inside of the downstream face of each light guide near the peak of the 
M{\o}ller flux, shown in the center frame of Figure~\ref{fig:MOLLER_detdiag}. 
For the scintillation background, the contribution is between 0.01 and 0.05 P.E.'s
depending on light guide
reflectivity. As noted in the benchmark simulation discussion earlier, the simulation and data differ from each other
by about 0.003 P.E. or about 40\%. As a conservative upper limit, we assume a reflectivity of 0.9 for the Cherenkov
light in our estimates; we expect the real design to virtually eliminate the Cherenkov component. 

The detector assembly shown in Figure~\ref{fig:MOLLER_detdiag} has been designed such
that the scattered flux is approximately normal to the fused silica tile and by
the simulated result shown in Figure \ref{fig:detector_sim_both}, in the region of minimum background. The
anticipated signal from the fused silica tile assuming a thickness of 1.5~cm is
about 45 P.E.'s per incident electron, nearly a factor of 1000 higher.
The ratio of the flux of charged particles in a tile to that in its corresponding
light guide varies dramatically and we consider here two important cases. 

In the first case we consider the ring that will detect the bulk of the M{\o}ller scattering signal and where background suppression is of paramount importance. We find that the relative light guide background rate ranges between 0.1 to 0.3\%\ depending on the azimuthal positioning of the
tile.  We expect to measure these background rates directly
in dedicated calibration runs and that the asymmetry of this background is much smaller than the M{\o}ller asymmetry such that we can control the 
extracted signal systematic error at the 0.1\%\ level. It is worth noting however that if the estimate were 3 to 6 times higher, as would be the case if one used dry N$_2$ as the medium, the background contamination would be uncomfortably high. 

In the second case we consider the ring which will predominantly detect the scattered electrons from
 inelastic electron-proton scattering, which will have light guide backgrounds dominated by the
large primary M{\o}ller scattered flux. The relative light yield from background and signal ranges from 5 to 20\% due to the relatively lower rate of the inelastic events. We would like to control the systematic error on the background corrections on this measurement at the 10\%\ level, which will be possible with this level of background. Again, if it were 3 to 6 times bigger, that would likely be unacceptably large. 

Based on the analysis above, we have concluded that light guides with air as the medium will be acceptable for the 
MOLLER apparatus.


\section{Summary}

A new gas-filled detector was designed to measure both the Cherenkov and scintillation response for high energy
electrons traversing different gases in order to quantify an important source of background for the detectors proposed in the MOLLER experiment.   The detector was deployed in a dedicated measurement using the electron beam at MAMI and the data were analyzed and used to validate a Geant4 simulation.  With the simulation, backgrounds due to Cherenkov and scintillation processes in air-filled light guides were estimated for the proposed design of the MOLLER detector
assembly and found to be manageable.  The Cherenkov background was also found to be strongly suppressed by 
judiciously blackening small sections of the light guide as well as found to have a minimum for a particular electron incident angle.

\section*{Acknowledgements}
We would like to thank Tom Hemmick and Craig Woody for useful suggestions and important contributions to the test apparatus, and the technical staff of the MAMI facility in Mainz, Germany. This work was supported by the U.S. Department of Energy and NSF in the United States, NSERC in Canada and DFG in Germany.


\bibliography{reference}

\begin{thebibliography}{1}
\expandafter\ifx\csname url\endcsname\relax
  \def\url#1{\texttt{#1}}\fi
\expandafter\ifx\csname urlprefix\endcsname\relax\def\urlprefix{URL }\fi
\expandafter\ifx\csname href\endcsname\relax
  \def\href#1#2{#2} \def\path#1{#1}\fi

\bibitem{Benesch:2014bas}
J.~Benesch, et~al., {The MOLLER Experiment: An Ultra-Precise Measurement of the
  Weak Mixing Angle Using M{\o}ller Scattering}\href
  {http://arxiv.org/abs/1411.4088} {\path{arXiv:1411.4088}}.

\bibitem{moller_homepage}
{MOLLER} experiment homepage, \url{http://hallaweb.jlab.org/12GeV/Moller/}.

\bibitem{moller_proposal}
{MOLLER}: Jefferson lab experiment {E12-09-005},
  \url{http://hallaweb.jlab.org/12GeV/Moller/pubs/moller_proposal.pdf}.

\bibitem{MORII2004399}
H.~Morii, K.~Mizouchi, T.~Nomura, N.~Sasao, T.~Sumida, M.~Kobayashi,
  Y.~Murayama, R.~Takashima, Quenching effects in nitrogen gas scintillation,
  Nuclear Instruments and Methods in Physics Research Section A: Accelerators,
  Spectrometers, Detectors and Associated Equipment 526~(3) (2004) 399 -- 408.

\bibitem{Bellamy:1994bv}
E.~H. Bellamy, G.~Bellettini, F.~Gervelli, M.~Incagli, D.~Lucchesi,
  C.~Pagliarone, F.~Zetti, {\relax Yu}.~Budagov, I.~Chirikov-Zorin, S.~Tokar,
  {Absolute calibration and monitoring of a spectrometric channel using a
  photomultiplier}, Nucl. Instrum. Meth. A339 (1994) 468--476.

\bibitem{AGOSTINELLI2003250}
Geant4-a simulation toolkit, Nuclear Instruments and Methods in Physics
  Research Section A: Accelerators, Spectrometers, Detectors and Associated
  Equipment 506~(3) (2003) 250 -- 303.

\bibitem{GERNHAUSER1996300}
R.~Gernhäuser, B.~Bauer, J.~Friese, J.~Homolka, A.~Kastenmüller, P.~Kienle,
  H.-J. Körner, P.~Maier-Komor, M.~Münch, R.~Schneider, K.~Zeitelhack, Photon
  detector performance and radiator scintillation in the hades rich, Nuclear
  Instruments and Methods in Physics Research Section A: Accelerators,
  Spectrometers, Detectors and Associated Equipment 371~(1) (1996) 300 -- 304.

\bibitem{rii}
M.~N. Polyanskiy, Refractive index database,
  \url{https://refractiveindex.info}.

\end{thebibliography}

\end{document}